\documentclass{IEEEtran}
\usepackage{amsmath}
\usepackage{nccmath}
\usepackage[utf8]{inputenc} 
\usepackage{epsfig,scalefnt,multirow}
\usepackage{url}
\usepackage{ifthen}
\usepackage{mathtools}
\usepackage{cite}
\usepackage{graphicx}
\usepackage{amssymb}
\usepackage{tabularx}
\usepackage{amsmath}
\usepackage{epstopdf}
\usepackage{epsf}
\usepackage{algorithm}
\usepackage{algpseudocode}
\usepackage{algpascal}
\usepackage{cases}
\usepackage{subfig}
\usepackage{stfloats}
\usepackage{float}
\usepackage{xcolor}
\usepackage{tabularx}

\usepackage{epsfig,amsmath,amssymb,epsf,amsthm,scalefnt,multirow,subfig}
\usepackage{xcolor}
\usepackage{float}
\usepackage{cite}
\usepackage{psfrag}
\usepackage{gensymb}
\usepackage{cleveref}
\usepackage{mathrsfs}
\usepackage{mathtools}
\usepackage{algpseudocode}
\usepackage{algorithm}
\usepackage{enumerate}

  \def\cC{{\mathcal{C}}}

 \def\cN{{\mathcal{N}}}

\def\argmax{\mathop{\mathrm{argmax}}}
\def\diag{\mathop{\mathrm{diag}}}

\def\bSigma{{\pmb{\Sigma}}} 
\def\bPhi{{\pmb{\Phi}}}

\def\b0{{\pmb{0}}} 

\def\ba{{\mathbf{a}}}   
  \def\bg{{\mathbf{g}}} \def\bh{{\mathbf{h}}}
   
\def\bm{{\mathbf{m}}}   
   
  \def\bw{{\mathbf{w}}}

\def\bA{{\mathbf{A}}}   
   \def\bH{{\mathbf{H}}}
\def\bI{{\mathbf{I}}}

  \def\bW{{\mathbf{W}}}

\DeclarePairedDelimiter\norm{\lVert}{\rVert}

\begin{document}
	%

	\title{Alternating Beamforming with \\Intelligent Reflecting Surface Element Allocation}

	%
	%
	%
	
	\author{Hyesang Cho,~\IEEEmembership{Student Member,~IEEE}	and~Junil Choi,~\IEEEmembership{Senior Member,~IEEE}
		\thanks{The authors are with the School of Electrical Engineering, Korea
			Advanced Institute of Science and Technology, Daejeon 34141, South Korea
			(e-mail: \{nanjohn96, junil\}@kaist.ac.kr).
	
	This research was supported by the MSIT (Ministry of Science and ICT), Korea, under the ITRC (Information Technology Research Center) support program (IITP-2020-0-01787) supervised by the IITP (Institute of Information \& Communications Technology Planning \& Evaluation) and the National Research Foundation (NRF) Grant funded by the MSIT of the Korea Government (2019R1C1C1003638).
	}
}

	\maketitle
	

\begin{abstract} \label{sec:abs} 
Intelligent reflecting surface (IRS) has become a promising technology to aid next generation wireless communication systems. In this paper, we develop an alternating beamforming technique with a novel concept of IRS element allocation for multiple-input multiple-output systems when a base station supports multiple single antenna users aided with a single IRS. Specifically, we allocate each IRS element separately to each user, thus, in the beamforming stage allowing the IRS element only consider a single user at a time. In result to this separation, the complexity is vastly decreased. The proposed beamforming technique tries to maximize the minimum rate of all users with minimal complexity. In the numerical results, we show that the proposed technique is comparable to the convex optimization-based benchmark with sufficiently less complexity. 
\end{abstract}






\section{Introduction}\label{sec:intro}
It has become the era of extravagant data where demands for wireless communication systems supporting extremely large data rates naturally follow. To satisfy this need, extensive research has been investigated through many attempts such as massive multiple-input multiple-output (MIMO) and millimeter wave communication \cite{Popo1, Tufv1, Schulz1}. However, practical issues such as power consumption and hardware cost are still in demand \cite{Schober1, Li1}.

Intelligent reflecting surface (IRS) is a promising novel technology having enormous potential, being a powerful candidate for future wireless communication systems\cite{Chen1,Bjorn1}. It is a planar array surface equipped with passive elements each having the ability to shift the phase of impinging electromagnetic waves when reflected \cite{Renzo,Aky1}. One core feature of IRS is that the phase shifts are independently controllable. It has the capability of changing wireless channel characteristics, e.g., a multipath profile, to increase data rates or get rid of shadowing effects.  Another main advantage of IRS is that by using passive elements, it is power- and cost-friendly.

Recently, there have been many works on IRS including beamforming, channel estimation, or physical channel modeling\cite{Zhang3, Poor1, Alouini1, Zhang5,Larsson2}. Most of works on IRS beamforming focused on alternating optimization, i.e., fixing the IRS phase shifts while optimizing the transmit beamformer at the base station and fixing the transmit beamformer when adjusting the IRS phase shifts. Especially, \cite{Alouini1} and \cite{Nallanathan1} proposed alternating optimization techniques with the objective functions of minimum rate maximization or multicast group sum rate maximization, respectively. Although effective, most of alternating optimization techniques are based on complex optimization methods and suffer from high computational complexity. Due to its cost-effectiveness, IRS is expected to have a large number of passive elements, thus, practical beamforming techniques must operate with low complexity.

 \begin{figure}[t] 
	\centering
	\includegraphics[width=0.85 \columnwidth]{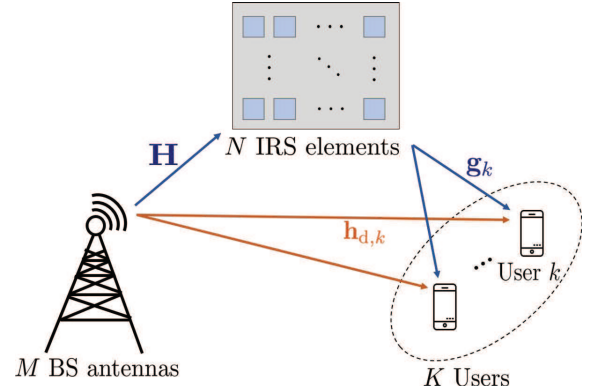}
	\caption{MU-MIMO downlink system of BS with $M$ antennas, $K$ UEs, and IRS with $N$ elements.} 
	\label{fig:System model}
\end{figure}


In this paper, we propose a low complexity beamforming technique with the purpose of satisfying one of the key purposes of IRS, supporting the users with poor channel conditions \cite{Zhang5}. To decrease complexity, we propose a novel concept of IRS element allocation. IRS element allocation is a selection technique similar to antenna sub-array formation, which groups transmit antennas into several subsets to transmit data streams with a limited number of RF chains \cite{Pi1}. The proposed IRS element allocation is based on the concept of allocating each IRS element to a single user, thus, allowing a single IRS element to consider only one user in the phase updating stage. Contrast to the conventional IRS beamforming techniques where each IRS element jointly considers all users simultaneously, the complexity of proposed beamforming technique is substantially reduced. Even with extremely low complexity, we confirm that the proposed technique only has marginal performance loss compared to the minimum rate maximization technique based on complicated optimization process in \cite{Alouini1}. To the best of our knowledge, this is the first attempt to use a grouping technique on a single IRS for a multiuser (MU) scenario, and to develop a suitable alternating beamforming technique to support this concept.

The rest of paper is organized as follows. Section \ref{sec:model} describes the system model of interest. Section \ref{sec:allocation} explains the proposed IRS element allocation and alternating beamforming technique, and Section \ref{sec: simul} shows simulation results of the proposed technique with several benchmark schemes. Finally, Section \ref{sec:conclu} concludes the paper. 

\textbf{Notation:} Lower and upper boldface letters represent column vectors and matrices. $\bA^{*}$  and $\bA^{\mathrm{H}}$ denotes the conjugate, and conjugate transpose of the matrix $\bA$. $\diag(\ba)$ returns the diagonal matrix with $\ba$ on its diagonal. ${\mathbb{C}}^{m \times n}$ and ${\mathbb{R}}^{m \times n}$ represent the set of all $m \times n$ complex and real matrices. $|{\cdot}|$ denotes the amplitude of the scalar, and $\norm{\cdot}$ represents the $\ell_2$-norm of the vector. $\lfloor a \rfloor$ denotes the integer less than or equal to the real number $a$.  $\mathcal{O}$ denotes the Big-O notation. $\boldsymbol{0}_m$ is used for the $m\times1$ all zero vector, and $\bI_m$ denotes the $m \times m$ identity matrix. $\cC \cN(\bm,\bSigma)$ denotes the circularly symmetric complex Gaussian distribution with mean $\bm$ and variance $\bSigma$.

\section{System Model}\label{sec:model}
In this paper, we focus on a single-cell MU-MIMO downlink system with one IRS as in Fig. \ref{fig:System model}. The base station (BS) with $M$ antennas is serving $K$ single-antenna user equipments (UEs) with the aid from the IRS with $N$ elements. As in \cite{Nallanathan1, Guo1}, we assume the BS has perfect channel state information for all links. The received signal at the $k$-th UE is denoted as
\begin{align}\label{eq:model}
y_k = \left( \bH \bPhi\bg_{k}+ \bh_{\mathrm{d},k}\right) ^\text{H} \sum_{i=1}^{K}\bw_is_i + z_k,
\end{align}
where $\bh_{\mathrm{d},k} \in \mathbb{C}^{M \times 1}$ is the direct channel from the BS to the $k$-th UE, $\bg_{k} = [g_{1,k}, g_{2,k}, ..., g_{N,k}]^{\mathrm{T}} \in \mathbb{C} ^{N \times 1}$ is the channel from the IRS to the $k$-th UE where $g_{n,k}$ is the channel between the $n$-th IRS element and the $k$-th UE, and $\bH = [\bh_1, \bh_2, ... ,\bh_N] \in \mathbb{C}^{M \times N}$ is the channel from the BS to the IRS where $\bh_n$ denotes the channel between the BS and the $n$-th IRS element. The transmit beamforming vector for the signal $s_i$ is $\bw_i \in \mathbb{C}^{M \times 1}$ with power constraint $\sum^K_{i=1}\norm{\bw_i}^2 \leq P$, and $z_k \sim  \mathcal{CN}(0, \sigma^2)$ is the white Gaussian noise. The phase shifts of the IRS elements are represented as
\begin{align}\label{eq:phase}
\bPhi &= \diag \left(\left[\phi_1, \phi_2, ..., \phi_N \right]\right), \\
\phi_n &= e^{j\theta_n}, \ \theta_n \in \left [0, 2\pi \right), \notag 
\end{align}
where $\phi_n$ is the reflection coefficient for the $n$-th IRS element.

With the received signal, the rate per UE is
\begin{align}\label{eq:SINR}
R_k &= B\log (1 + \gamma_k), \notag \\
\gamma_k &= \frac{\left| \left( \bH \bPhi\bg_{k}+ \bh_{\mathrm{d},k}\right)^\text{H}\bw_k\right|^2}{\sum^K_{i \ne k}\left| \left(\bH \bPhi\bg_{k}+ \bh_{\mathrm{d},k}\right)^\text{H}\bw_i\right|^2  + \sigma^2B},
\end{align}
where $\gamma_k$ is the signal-to-interference-and-noise ratio (SINR) for the $k$-th UE, and $B$ is the system bandwidth. The minimum rate (min-rate) is defined as 
\begin{align}
	R_{\mathrm{min}} = \min_k (R_1, ... , R_K),
\end{align}
which is the rate of the UE with the smallest achievable rate.

\section{Proposed Alternating Beamforming with IRS Element Allocation}\label{sec:allocation}
In this section, we first discuss the novel IRS element allocation technique, followed by the alternating beamforming reinforced by the IRS element allocation. Then, we illustrate the rationale of our proposed algorithm and summarize all the steps to explicitly develop the proposed min-rate maximization beamforming algorithm.

\subsection{IRS Element Allocation}\label{subsec:IRS allocation}
The proposed IRS element allocation is based on the key idea of matching one IRS element to one UE to reduce complexity. However, this concept should be more specified with two factors: 1) the number of IRS elements each UE should be assigned to, and 2) the selection of specific IRS elements each UE should be allocated with. These factors depend on our main objective, i.e., maximizing the minimum rate. 
To achieve this goal, we assign more IRS elements to the weak UEs. In mathematical sense, 
\begin{align} \label{eq:assign}
\alpha_k =  \frac{1}{p_k}, \quad 
\ell_k = \left\lfloor N\frac{\alpha_k}{\sum_{i=1}^{K}\alpha_i} \right\rfloor,
\end{align}
where $p_k$ is the metric for assigning the IRS elements, $\frac{\alpha_k}{\sum_{i=1}^{K}\alpha_i}$ is the proportion of IRS elements the $k$-th UE is assigned, and $\ell_k$ is the number of IRS elements assigned to the $k$-th UE. Note that $p_k$ is also a design parameter. Although straightforward metrics are $R_k$ or $\gamma_k$, there are also indirect metrics such as $\left| \bh_{\mathrm{d},k}^\text{H}\bw_k\right|$, i.e., the direct channel beamforming gain. We have verified that the metric $\left| \bh_{\mathrm{d},k}^\text{H}\bw_k\right|$ gives performance similar to that using other metrics $R_k$ and $\gamma_k$, although these results are not included in the paper due to limited space. Therefore, we set $p_k = \left| \bh_{\mathrm{d},k}^\text{H}\bw_k\right|$ henceforth. The remaining IRS elements are assigned to the weakest UE as
\begin{align} \label{eq:assign2}
	r &= N - \sum_{k=1}^{K}\ell_k,	\\
	\ell_{k_0} &= \ell_{k_0}+r, \ k_0 = \argmax_k \alpha_k. \notag
\end{align}
Note that the proposed allocation method does not assign all the IRS elements to the weakest UE but rather it assigns the elements to the UEs inversely proportional to the direct channel beamforming gains. This is to make all the UEs have a decent channel condition by using the IRS. 


Although the number of IRS elements allocated to each UE is determined, still we need to define which IRS elements to allocate to each UE. Note that the main purpose of proposed method is to help the weak UEs more. Thus, we sort the UEs in descending order with respect to $\alpha_k$ as
\begin{align}
\alpha_{m_1} \geq \alpha_{m_2} \geq ... \geq \alpha_{m_K}.
\end{align}
Then, starting from the $m_1$-th UE, we allocate the strongest IRS elements with respect to the selected UE, e.g., for the $m_1$-th UE, we allocate $\ell_{m_1}$ IRS elements with the strongest channel beamforming gains within $\bg_{m_1}^{\text{H}}\bH^{\text{H}} \bw_{m_1}$, i.e., the BS-IRS-UE channel beamforming gain. By allocating $\ell_{m_i}$ IRS elements to the $m_i$-{th} UE, each IRS element will be allocated to only one UE. Since this allocation process is mixed with the proposed alternating beamforming technique, we leave the mathematical definition of allocation process to Section \ref{sec: overall}.

\subsection{Alternating Beamforming Technique} \label{sec: alternating beam}
In this subsection, we illustrate the proposed transmit beamformer and phase updating technique. Similar to \cite{Zhang2, Alouini1}, we alternatively update the IRS phases and the transmit beamformer at the BS, but with the additional stage of IRS element allocation. The proposed alternating beamforming technique, however, does not rely on any complex optimization process for all three stages, vastly decreasing the complexity compared to the previous works.

We use the superscript $v$ to denote the $v$-th update of beamformer and IRS phases, i.e., $\bW^v = [\bw_1^v, ... , \bw_K^v]$, where $\bw_k^v$ is the $v$-th transmit beamformer update for the $k$-th UE, and $\bPhi^v$ is the $v$-th IRS phase update as $\bPhi^v = \diag\left(\left[\phi^v_1,\phi^v_2, ...,\phi^v_N\right]\right)$  with $\phi^v_n = e^{j\theta_n^v}$ the phase of the $n$-th IRS element.

\subsubsection{Transmit Beamformer Update}
Recalling the system model, the received signal at the $k$-th UE with the $v$-th iteration of transmit beamformer and IRS phase update can be shown as 
\begin{align}
y_k = \left( \bH \bPhi^v\bg_{k}+ \bh_{\mathrm{d},k}\right) ^\text{H} \sum_{i=1}^{K}\bw_i^vs_i + z_k.
\end{align}
By treating $\left( \bH \bPhi^v\bg_{k}+ \bh_{\mathrm{d},k}\right)$ as an effective channel, the overall channel boils down to the conventional MU-MIMO downlink channel. Therefore, we may use well-known transmit beamformers such as the zeroforcing (ZF) beamformer, regularized ZF (RZF) beamformer, or the maximum ratio transmission (MRT) beamformer. Note that we also have the flexibility to choose different beamformers at each iteration to our benefit. In Section \ref{sec: simul}, we numerically verify the effect of each transmit beamformer on the proposed technique.

\subsubsection{IRS Phase Update}
With the fixed transmit beamformer at the BS, the SINR of the $k$-th UE is given as \eqref{eq:SINR}. By expanding $\bH$ and $\bg_{k}$, $\gamma_k$ can be reformulated as
\begin{align} \label{eq:alternate sinr}
\gamma_k =\frac{\left|\bh_{\mathrm{d},k}^\text{H}\bw_{k}^{v} +\sum_{n=1}^{N} g_{n,k}^*\left(\phi_n^{v}\right)^*\bh_n^\text{H}\bw_{k}^{v}\right|^2}{\sum_{i \ne k}^{K}\left|\bh_{\mathrm{d},k}^\text{H}\bw_{i}^{v} +\sum_{n=1}^{N} g_{n,k}^*\left(\phi_n^{v}\right)^*\bh_n^\text{H}\bw_{i}^{v}\right|^2+ \sigma^2B},
\end{align}
with the $v$-th update of transmit beamformer and IRS phases. Considering only the numerator in \eqref{eq:alternate sinr}, we can use the triangle inequality to find the upper-bound of the numerator,
\begin{align} \label{eq:phase rationale}
	&\left|\bh_{\mathrm{d},k}^\text{H}\bw_{k}^{v} +\sum_{n=1}^{N} g_{n,k}^*\left(\phi_n^v\right)^*\bh_n^\text{H}\bw_{k}^{v}\right| \notag \\ 
	 &\quad \quad \quad \quad \quad \leq \left|\bh_{\mathrm{d},k}^\text{H}\bw_{k}^{v}\right| +  \sum_{n=1}^{N}\left| g_{n,k}^*\bh_n^\text{H}\bw_{k}^{v}\right|,
\end{align}
where the equality holds when all the phases of IRS elements are properly aligned such that the $k$-th UE receives coherently combined signal. It may not be possible, however, to achieve the equality condition for all UEs since the IRS phases affect all the UEs simultaneously.

Assuming the IRS element allocation stage is finished, we align the BS-IRS-UE channels to the direct channel of the UEs, i.e., the phase of $n$-th IRS element assigned to the $k$-th UE is set as
\begin{align} \label{eq: phase update}
	\angle \theta_n^{v+1} = -\angle(\bh_{\mathrm{d},k}^\text{H}\bw_{k}^{v}) -\angle g_{n,k} + \angle(\bh_n^\text{H}\bw_{k}^{v}), \ n \in \cN_k,
\end{align}
which makes
\begin{align}
\angle \left(\bh_{\mathrm{d},k}^\text{H}\bw_{k}^{v}\right)=\angle \left(g_{n,k}^*\left(\phi_n^{v+1}\right)^*\bh_n^\text{H}\bw_{k}^{v} \right), \ n \in \cN_k,
\end{align}
where $\cN_k$ denotes the set of IRS elements allocated to the $k$-th UE. After updating all IRS elements, we can express the numerator in \eqref{eq:alternate sinr} as 
\begin{align} \label{numerator}
	\left(\left|\bh_{\mathrm{d},k}^\text{H}\bw_{k}^{v}\right| + \sum_{n \in \cN_k}\left|g_{n,k}^*\bh_n^\text{H}\bw_{k}^{v}\right|\right)e^{j\angle \left(\bh_{\mathrm{d},k}^\text{H}\bw_{k}^{v}\right)}+ o_k^{v+1},
\end{align}
where $o_k^{v+1}$, which represents the combined signals from the IRS elements allocated to different UEs, is given as 
\begin{align}
	o_k^{v+1} = \sum_{n \notin \cN_k}g_{n,k}^*\left(\phi_n^{v+1}\right)^*\bh_n^\text{H}\bw_{k}^{v}.
\end{align}
Due to the fact that each IRS element considers only one UE, each IRS element phase update can be expressed in closed form as in \eqref{eq: phase update}. Thus, by coherently combining the signal to a specific UE rather than jointly considering all UEs, the phase update can be computed with minimal complexity.

Note that in the IRS phase update process, we only focused on the received signal part without considering the interference, i.e., the denominator of 
\eqref{eq:alternate sinr}. Although simple, the proposed IRS phase update works quite well as explained in the next subsection.

\subsection{Rationale of Proposed Alternating Beamforming}
In the transmit beamformer updating stage, recall that the effective channel is the conventional downlink channel. As the transmit beamformer and IRS phases are alternatively updated, the IRS phase update will change the effective channel, making the previous transmit beamformer inadequate to the new effective channel. For instance, let us assume that we use the equal power MRT beamformer. Then, in the transmit beamformer update,
\begin{align} \label{beam upate}
	\bh_{\mathrm{eff},k}^{v} &= \bH \bPhi^{v}\bg_{k}+ \bh_{\mathrm{d},k}, \notag \\
	\bw^{v}_k &= \frac{P}{K}\frac{\bh^v_{\mathrm{eff},k}}{\norm{\bh^v_{\mathrm{eff},k}}}.
\end{align}
It is highly likely that $\bW^{v-1}$ is outdated for $\bh_{\mathrm{eff},k}^{v}, \ \forall k \in \left\{1, 2, ..., K\right\}$ due to the $v$-th IRS phase update. Thus, by updating the transmit beamformer, with high probability the performance, in this case the min-rate, would increase.

In the IRS element allocation stage, the IRS elements will be allocated to the UEs according to Section \ref{subsec:IRS allocation}. Note that as the transmit beamformer is updated, the beamforming gain of the UEs are also updated. Thus, by updating the IRS element allocation, the number of IRS elements will be newly assigned followed by newly allocating the IRS elements. This stage will prevent the situation of reinforcing wrong UEs or allocating inadequate IRS elements during iterations.

In the IRS phase updating stage, $\gamma_k$ with $\bW^v$ and $\bPhi^v$ is given as
\begin{align}
\gamma_k =\frac{\left|\bh_{\mathrm{d},k}^\text{H}\bw_{k}^v +\sum_{n=1}^{N} g_{n,k}^*\left(\phi_n^v\right)^*\bh_n^\text{H}\bw_{k}^v\right|^2}{\sum_{i \ne k}^{K}\left|\bh_{\mathrm{d},k}^\text{H}\bw_{i}^v +\sum_{n=1}^{N} g_{n,k}^*\left(\phi_n^v\right)^*\bh_n^\text{H}\bw_{i}^v\right|^2+ \sigma^2B}.
\end{align}
Although not explicitly shown, the phases of BS-IRS-UE channels may not be aligned to the direct paths due to the transmit beamformer update. After the IRS phase update, $\gamma_k$ is given as
\begin{align} \label{SINR update}
\gamma_k =\frac{\left(\left|\bh_{\mathrm{d},k}^\text{H}\bw_{k}^{v}\right| + \sum_{n \in \cN_k}\left|g_{n,k}^*\bh_n^\text{H}\bw_{k}^{v}\right|\right)e^{j\angle \left(\bh_{\mathrm{d},k}^\text{H}\bw_{k}^{v}\right)} + o_k^{v+1}}{\sum_{i \ne k}^{K}\left|\bh_{\mathrm{d},k}^\text{H}\bw_{i}^v +\sum_{n=1}^{N} g_{n,k}^*\left(\phi_n^{v+1}\right)^*\bh_n^\text{H}\bw_{i}^v\right|^2+ \sigma^2B}.
\end{align}
In \eqref{SINR update}, since the IRS phases for the $k$-th UE are uniquely determined by $\bw_k$, the BS-IRS-UE channels for $n \in \cN_k$ add up constructively in the numerator, but not in the denominator. Thus, by updating all the IRS elements, the rate of all UEs would increase with high probability. Due to the interference, the performance increase would not be for each instance though.

\subsection{Summary of Proposed Beamforming Technique and Complexity Discussion} \label{sec: overall}
The overall proposed alternating beamforming technique is summarized in Algorithm 1 where $\cN_0$ is the set of unallocated IRS elements. First, with the transmit beamformer fixed, assign the IRS elements to the UEs. Then, by referring to \eqref{eq:assign2} and \eqref{eq:phase rationale}, we allocate $\ell_{m_i}$ IRS elements to the $m_i$-th UE using the BS-IRS-UE channel beamforming gain $\left|g_{n,k}^*\bh_n^{\text{H}}\bw_{m_i}\right|$ as the metric. Thus, the allocation of a single IRS element can be shown as
\begin{align}
	n_0 = \argmax_n \left| g_{n,m_i}^*\bh_n^{\text{H}}\bw_{m_i}\right|, \ n \in \cN_0, \notag \\
	\cN_{m_i} = \cN_{m_i} + \{ n_0\}, \ \cN_0 = \cN_0 - \{ n_0\}.
\end{align}
The IRS element phase update and transmit beamformer update are then performed after allocating the IRS elements. By repeating this sequence iteratively $V$ times, we get the results.

\begin{algorithm}[t]
	\begin{algorithmic} [1]
		\caption{Pseudo-code for IRS allocation and alternating beamforming}
		\State Initialization: Random $\bPhi, \bW$, $\cN_0 = \{1, 2, ...,N\}, \cN_k = \emptyset, \ 1 \leq k \leq K$
		\Repeat
		\State With $\alpha_k =  \frac{1}{\left| \bh_{\mathrm{d},k}^\text{H}\bw_k\right|}$, set $\ell_k = \left\lfloor N\frac{\alpha_k}{\sum_{i=1}^{K}\alpha_i} \right\rfloor$ with remainders as \eqref{eq:assign2}
		\State Order UEs as $\alpha_{m_1} \geq \alpha_{m_2} \geq ... \geq \alpha_{m_K}$
		\For{i=1}{K}{}
		\Repeat
		\State $n_0 = \argmax_n |g_{n,m_i}^*\bh_n^{\text{H}}\bw_{m_i}|, \ n \in \cN_0$
		\State $\cN_{m_i} = \cN_{m_i} + \{ n_0\}, \ \cN_0 = \cN_0 - \{ n_0\}$
		\Until \ Iteration is repeated $\ell_{m_i}$ number of times
		\EndFor
		\For{k=1}{K}{}
		\Repeat
		\State $\angle \theta_{n_k} = -\angle(\bh_{\mathrm{d},k}^\text{H}\bw_{k}) -\angle g_{n_k,k} + \angle(\bh_{n_k}^\text{H}\bw_{k}),$
		\State $\cN_k = \cN_k - \{ n_k\}, \ \cN_0 = \cN_0 + \{ n_k\}$
		\Until \ $\cN_k = \emptyset $
		\State Update transmit beamformer $\bW$ using effective channel formed by $\bPhi$
		\Until \ Iteration is repeated $V >0$ number of times
	\end{algorithmic}
\end{algorithm} 


Note that with the flexibility of the transmit beamforming stage, we can use various types of transmit beamformers including the min-rate maximizing beamformer \cite{Tan1}, or RZF, ZF, MRT beamformers to decrease complexity with possible performance trade-off. If we use equal power RZF as the intermediate transmit beamformer for each iteration and the min-rate maximizing beamformer for the final transmit beamformer update, the complexity of the proposed algorithm is of the order $\mathcal{O} \left( (V+q)M^3K\right)$ when \textit{N}$N$ is small, and $\mathcal{O} \left( V(M+1)N(N+1)\right)$ when $N$ is large, where $q$ is the number of iterations for the min-rate maximizing beamformer. Note that the benchmarks we adopt in the simulations, i.e., \cite{Alouini1,Guo1}, have the maximum complexity of order $\mathcal{O} \left( VN^{4.5}\right)$ \cite{Zhang4}, which is excessive, especially when $N$ is large. Due to the fact that IRS will have a large number of elements compared to the numbers of BS antennas or UEs, the proposed algorithm is shown to have a drastic decrease in complexity.


\section{Simulation Results} \label{sec: simul}
In this section, we verify the proposed beamforming technique through Monte-Carlo simulations. The number of BS antennas, IRS elements, and UEs are $M = 8, N = 100,$ and $K = 4$ unless stated otherwise. The distance between the BS-IRS, BS-UE, and IRS-UE are $100$ m, $105$ m, and $10$ m, respectively. The pathloss model is defined as $\epsilon = \sqrt{d^{-\beta}}$ where $d$ is distance and $\beta$ is the pathloss exponent. We assume independent Rayleigh fading for all channels as $\bh_n = \epsilon_\text{BI} \boldsymbol{\eta}_N$, $\bh_{\mathrm{d},k} = \epsilon_\text{BU} \boldsymbol{\eta}_M$, and $\bg_k = \epsilon_\text{IU} \boldsymbol{\eta}_M$ where $\boldsymbol{\eta}_x \sim \cC\cN \left(\boldsymbol{0}_x, \bI_x \right)$. The pathloss exponent $\beta$ is $3.6, 4, 4$  for the BS-IRS, BS-UE, and IRS-UE. The noise power is determined with the noise power density $\sigma^2 = -174$ dBm/Hz and the bandwidth of $B = 10$ MHz for every UE.

Simulations are also held for cases of no IRS and with an IRS but only performing downlink transmit beamforming with random IRS phases. These two cases are denoted as ``No-IRS" and ``Random" in the following figures. There are two benchmarks proposed in \cite{Alouini1} and \cite{Guo1}, where \cite{Alouini1} exploits complex convex optimization techniques such as SDR, and \cite{Guo1} performs minimum secrecy rate maximization. For \cite{Guo1}, we modified the proposed algorithm by setting the rate of the eavesdroppers as zero and use the min-rate maximizing beamformer\cite{Tan1} for the last transmit beamformer. The number of iterations of the proposed algorithm is $V = 5$ for all cases. For the intermediate transmit beamformer of proposed algorithm, we take MRT, ZF, and RZF into account while the last transmit beamformer update is based on the min-rate maximizing beamformer \cite{Tan1}. We also use the same min-rate maximizing beamformer for baselines No-IRS and Random.

\begin{figure}[t]
	\centering
	\includegraphics[width=0.73 \columnwidth]{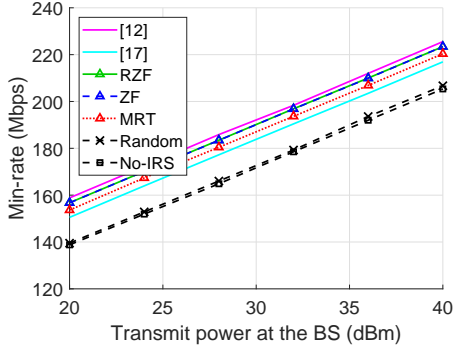}
	\caption{Min-rate with respect to the transmit power at the BS.}
	\label{fig:fix dist Power}
\end{figure}
In Fig. \ref{fig:fix dist Power}, we observe that the proposed technique with all three beamformers outperform No-IRS, Random, and the benchmark modified from \cite{Guo1}. Especially, the ZF beamformer and RZF beamformer have performance close to the SDR technique based benchmark, which is denoted as \cite{Alouini1}. The ZF and RZF beamformers are expected to outperform the MRT beamformer since they handle the interference in the transmit beamforming stage.

\begin{figure}[t]
	\centering
	\includegraphics[width=0.73 \columnwidth]{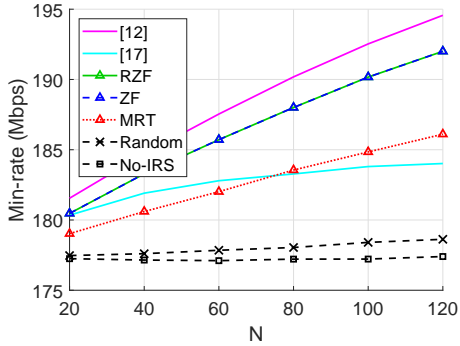}
	\caption{Min-rate with respect to the number of IRS elements $N$. Transmit power at the BS $P$ is fixed as $30$ dBm.}
	\label{fig:with N}
\end{figure}

In Fig. \ref{fig:with N}, we plot the min-rates according to the number of IRS elements. The figure shows that the performance of proposed technique increases faster than [17] with respect to the number of IRS elements, verifying that the proposed technique exploits the IRS elements well. However, we can also observe that the gap between\cite{Alouini1} and the proposed technique increases. This is due to the fact that each IRS element of proposed technique only focuses on one UE to reduce the complexity, thus, there is inevitable loss compared to \cite{Alouini1}, which updates each IRS element considering all UEs jointly. The joint update, however, is impractical due to excessive computational complexity. In fact, due to the complexity of\cite{Alouini1}, it is hard to compare the techniques in Fig. 3 with more IRS elements than 120.


\section{Conclusion} \label{sec:conclu}
In this paper, we proposed a novel alternating beamforming technique in a MU-MIMO downlink scenario with a BS serving multiple single antenna users aided with a single IRS. Through the novel concept of IRS element allocation and low-complexity alternative beamforming technique, the proposed beamforming technique enjoys very low-complexity while its performance is comparable to the benchmark based on the complex SDR process.


\bibliographystyle{IEEEtran}
\bibliography{Hyebib}

\begin{thebibliography}{10}
\providecommand{\url}[1]{#1}
\csname url@samestyle\endcsname
\providecommand{\newblock}{\relax}
\providecommand{\bibinfo}[2]{#2}
\providecommand{\BIBentrySTDinterwordspacing}{\spaceskip=0pt\relax}
\providecommand{\BIBentryALTinterwordstretchfactor}{4}
\providecommand{\BIBentryALTinterwordspacing}{\spaceskip=\fontdimen2\font plus
\BIBentryALTinterwordstretchfactor\fontdimen3\font minus
  \fontdimen4\font\relax}
\providecommand{\BIBforeignlanguage}[2]{{%
\expandafter\ifx\csname l@#1\endcsname\relax
\typeout{** WARNING: IEEEtran.bst: No hyphenation pattern has been}%
\typeout{** loaded for the language `#1'. Using the pattern for}%
\typeout{** the default language instead.}%
\else
\language=\csname l@#1\endcsname
\fi
#2}}
\providecommand{\BIBdecl}{\relax}
\BIBdecl

\bibitem{Popo1}
F.~{Boccardi}, R.~W. {Heath}, A.~{Lozano}, T.~L. {Marzetta}, and P.~{Popovski},
  ``{Five Disruptive Technology Directions for 5G},'' \emph{IEEE Communications
  Magazine}, vol.~52, no.~2, pp. 74--80, Feb. 2014.

\bibitem{Tufv1}
F.~{Rusek} \emph{et~al.}, ``{Scaling Up MIMO: Opportunities and Challenges with
  Very Large Arrays},'' \emph{IEEE Signal Processing Magazine}, vol.~30, no.~1,
  pp. 40--60, Dec. 2013.

\bibitem{Schulz1}
T.~S. {Rappaport} \emph{et~al.}, ``{Millimeter Wave Mobile Communications for
  5G Cellular: It Will Work!}'' \emph{IEEE Access}, vol.~1, pp. 335--349, May
  2013.

\bibitem{Schober1}
Q.~{Wu}, G.~Y. {Li}, W.~{Chen}, D.~W.~K. {Ng}, and R.~{Schober}, ``{An Overview
  of Sustainable Green 5G Networks},'' \emph{IEEE Wireless Communications},
  vol.~24, no.~4, pp. 72--80, Aug. 2017.

\bibitem{Li1}
S.~{Zhang}, Q.~{Wu}, S.~{Xu}, and G.~Y. {Li}, ``{Fundamental Green Tradeoffs:
  Progresses, Challenges, and Impacts on 5G Networks},'' \emph{IEEE
  Communications Surveys Tutorials}, vol.~19, no.~1, pp. 33--56, Jul. 2017.

\bibitem{Chen1}
W.~{Saad}, M.~{Bennis}, and M.~{Chen}, ``{A Vision of 6G Wireless Systems:
  Applications, Trends, Technologies, and Open Research Problems},'' \emph{IEEE
  Network}, vol.~34, no.~3, pp. 134--142, Oct. 2020.

\bibitem{Bjorn1}
E.~Björnson \emph{et~al.}, ``{Massive MIMO is a Reality—What is Next?: Five
  Promising Research Directions for Antenna Arrays},'' \emph{Digital Signal
  Processing}, vol.~94, pp. 3 -- 20, Nov. 2019.

\bibitem{Renzo}
M.~{Renzo}, M.~{Debbah}, and D.~{Phan-Huy}, ``{Smart Radio Environments
  Empowered by Reconfigurable AI Meta-Surfaces: An idea whose time has come.}''
  \emph{EURASIP Journal on Wireless Com Network}, no. 129, May 2019.

\bibitem{Aky1}
C.~{Liaskos} \emph{et~al.}, ``{A New Wireless Communication Paradigm Through
  Software-Controlled Metasurfaces},'' \emph{IEEE Communications Magazine},
  vol.~56, no.~9, pp. 162--169, Sep. 2018.

\bibitem{Zhang3}
E.~{Basar} \emph{et~al.}, ``{Wireless Communications Through Reconfigurable
  Intelligent Surfaces},'' \emph{IEEE Access}, vol.~7, pp. 116\,753--116\,773,
  Aug. 2019.

\bibitem{Poor1}
M.~Najafi, V.~Jamali, R.~Schober, and V.~H. Poor, ``{Physics-based Modeling and
  Scalable Optimization of Large Intelligent Reflecting Surfaces},''
  \emph{arXiv:2004.12957}, Apr. 2020.

\bibitem{Alouini1}
Q.~{Nadeem}, H.~{Alwazani}, A.~{Kammoun}, A.~{Chaaban}, M.~{Debbah}, and
  M.~{Alouini}, ``{Intelligent Reflecting Surface-Assisted Multi-User MISO
  Communication: Channel Estimation and Beamforming Design},'' \emph{IEEE Open
  Journal of the Communications Society}, vol.~1, pp. 661--680, May 2020.

\bibitem{Zhang5}
Q.~Wu, S.~Zhang, B.~Zheng, C.~You, and R.~Zhang, ``{Intelligent Reflecting
  Surface Aided Wireless Communications: A Tutorial},''
  \emph{arXiv:2007.02759}, Jul. 2020.

\bibitem{Larsson2}
H.~{Guo}, Y.~{Liang}, J.~{Chen}, and E.~G. {Larsson}, ``{Weighted Sum-Rate
  Maximization for Reconfigurable Intelligent Surface Aided Wireless
  Networks},'' \emph{IEEE Transactions on Wireless Communications}, vol.~19,
  no.~5, pp. 3064--3076, Feb. 2020.

\bibitem{Nallanathan1}
G.~{Zhou}, C.~{Pan}, H.~{Ren}, K.~{Wang}, and A.~{Nallanathan}, ``{Intelligent
  Reflecting Surface Aided Multigroup Multicast MISO Communication Systems},''
  \emph{IEEE Transactions on Signal Processing}, vol.~68, pp. 3236--3251, Apr.
  2020.

\bibitem{Pi1}
O.~E. {Ayach}, R.~W. {Heath}, S.~{Rajagopal}, and Z.~{Pi}, ``{Multimode
  Precoding in Millimeter wave MIMO Transmitters with Multiple Antenna
  Sub-arrays},'' in \emph{2013 IEEE Global Communications Conference}, Dec.
  2013, pp. 3476--3480.

\bibitem{Guo1}
J.~{Chen}, Y.~{Liang}, Y.~{Pei}, and H.~{Guo}, ``{Intelligent Reflecting
  Surface: A Programmable Wireless Environment for Physical Layer Security},''
  \emph{IEEE Access}, vol.~7, pp. 82\,599--82\,612, Jun. 2019.

\bibitem{Zhang2}
Q.~{Wu} and R.~{Zhang}, ``{Intelligent Reflecting Surface Enhanced Wireless
  Network via Joint Active and Passive Beamforming},'' \emph{IEEE Transactions
  on Wireless Communications}, vol.~18, no.~11, pp. 5394--5409, Aug. 2019.

\bibitem{Tan1}
D.~W.~H. {Cai}, T.~Q.~S. {Quek}, and C.~W. {Tan}, ``{A Unified Analysis of
  Max-Min Weighted SINR for MIMO Downlink System},'' \emph{IEEE Transactions on
  Signal Processing}, vol.~59, no.~8, pp. 3850--3862, May 2011.

\bibitem{Zhang4}
Z.~{Luo} \emph{et~al.}, ``{Semidefinite Relaxation of Quadratic Optimization
  Problems},'' \emph{IEEE Signal Processing Magazine}, vol.~27, no.~3, pp.
  20--34, Apr. 2010.

\end{thebibliography}

\end{document}